# Title: Differentiable rotamer sampling with molecular force fields

Authors: Congzhou M. Sha[1,2], Jian Wang[2], Nikolay V. Dokholyan[1,2,3,4,5,*]

Affiliations: [1]Department of Engineering Science and Mechanics, Penn State University, University Park, PA USA
[2]Department of Pharmacology, Penn State College of Medicine, Hershey, PA USA
[3]Department of Biochemistry and Molecular Biology, Penn State College of Medicine, Hershey, PA USA
[4]Department of Chemistry, Penn State University, University Park, PA USA
[5]Department of Biomedical Engineering, Penn State University, University Park, PA USA
*Corresponding author (dokh@psu.edu)

Abbreviations: molecular dynamics (MD), discrete molecular dynamics (DMD), Markov chain Monte Carlo (MCMC), Protein Data Bank (PDB), Kullback-Leibler divergence divergence (KL divergence), root mean square distance (RMSD)

**Graphical Abstract**

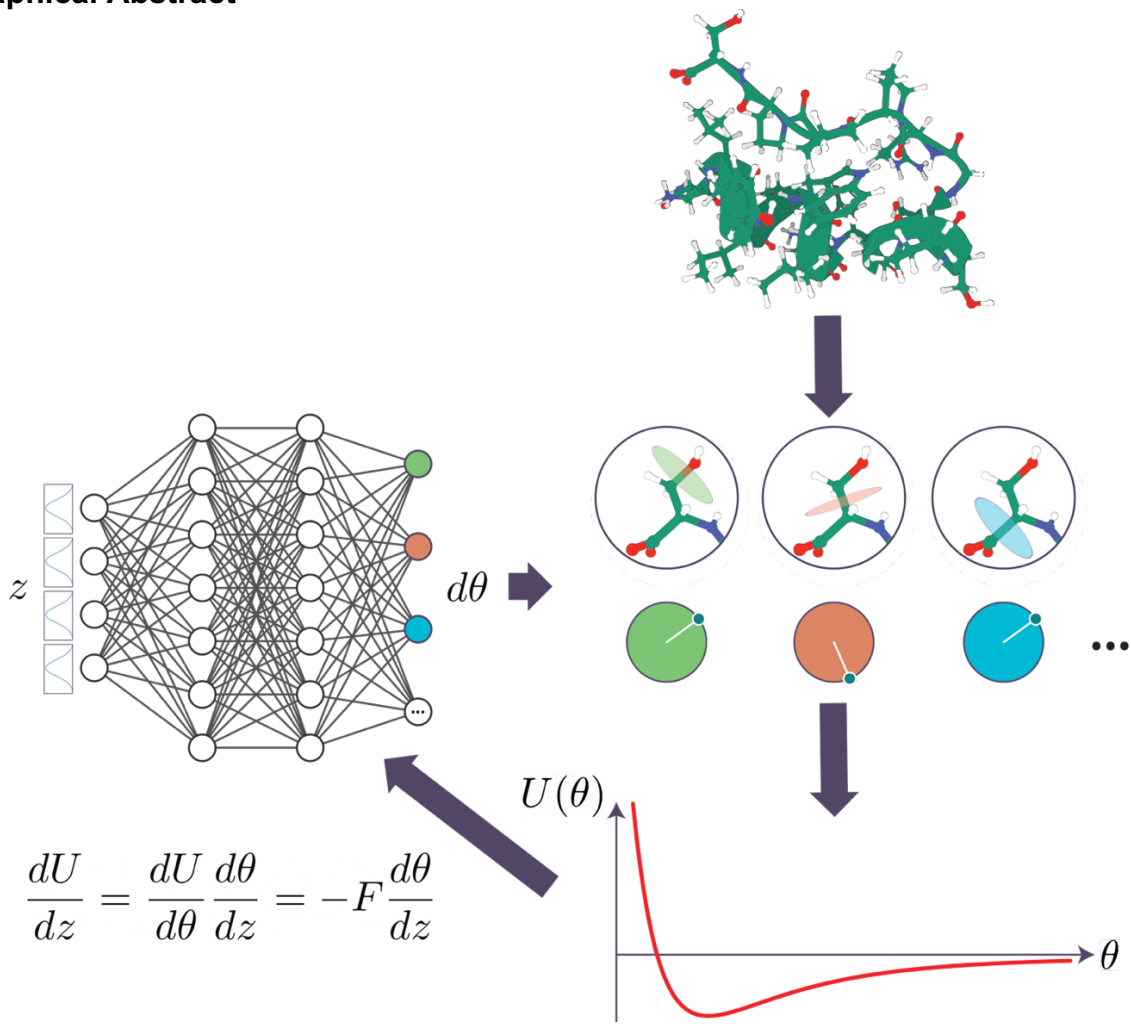

**One Sentence Summary:**

We make theoretical and practical improvements to the Boltzmann generator framework, enabling rapid and accurate sampling of macromolecule rotameric states and replacing molecular dynamics with neural network training.

**Abstract:**
Molecular dynamics is the primary computational method by which modern structural biology explores macromolecule structure and function. Boltzmann generators have been proposed as an alternative to molecular dynamics, by replacing the integration of molecular systems over time with the training of generative neural networks. This neural network approach to MD samples rare events at a higher rate than traditional MD, however critical gaps in the theory and computational feasibility of Boltzmann generators significantly reduce their usability. Here, we develop a mathematical foundation to overcome these barriers; we demonstrate that the Boltzmann generator approach is sufficiently rapid to replace traditional MD for complex macromolecules, such as proteins in specific applications, and we provide a comprehensive toolkit for the exploration of molecular energy landscapes with neural networks.

**Main Text:**

**Introduction**

Statistical mechanics describes the behavior of large numbers of physically identical systems (*1*). Molecular dynamics (MD) is the computational application of statistical mechanics to molecular systems such as proteins, nucleic acids, and lipid membranes (*2–5*). The fundamental postulate of statistical mechanics is that every energetically accessible microstate of the physical system is equally probable; a microstate is a partition of the total energy of the physical system to each coordinate of its Hamiltonian (*1*). When many identical copies of the physical system are present such as in molecular systems at equilibrium, experimental observations reflect the overall probability distribution of microstates.

The goal of MD is to computationally sample enough microstates of a system of molecules to approximate the distribution of microstates in a biological system at equilibrium, in which there may be on the order of Avogadro's number ($N_A \sim 10^{23}$) molecules. For MD, statistical equilibrium is defined as the NPT ensemble (*6*), in which the number of particles, pressure, and temperature are fixed, and the underlying microstate probability distribution is the Boltzmann distribution for the enthalpy.

Traditional MD attempts to sample microstates by integrating Newton's second law according to empirically determined molecular force fields (*2, 3*). The underlying major assumption is that the MD trajectory is *ergodic* (*6*), that is given enough time steps the trajectory will visit all microstates with a frequency given by the Boltzmann distribution. However, there is no guarantee that a given MD trajectory will be ergodic. Transitioning between states which are separated by large energy barriers presents a significant challenge for MD simulations (*7*). Numerous approaches have been proposed to address this shortcoming of MD, such as Monte Carlo methods (*8–11*), metadynamics (*12*), and umbrella sampling (*13*). Recently, Boltzmann generators have emerged as a promising candidate for rare-event sampling in MD (*14, 15*).

In 2019, foundational methods were proposed by Noé et al. to use generative neural networks for the sampling of microstates (*14*). The central idea is that instead of predicting a single trajectory as in MD, one may instead train a neural network to predict Boltzmann-distributed states. This approach seeks to address the issue of rare-event sampling by allowing the neural network to learn multiple energy minima simultaneously. While Noé et al. successfully demonstrated their method for simple physical systems and small proteins, there were critical theoretical and practical deficiencies limiting the application of their methods that we address in this work.

The theoretical deficiencies in the original Boltzmann generator approach we address are various biases in angle generation due to (i) the use of a Gaussian prior, and (ii) the regularization of a discontinuous output. The practical deficiencies we address are (iii) tight coupling between energy and entropy estimation, necessitating hundreds of thousands of evaluations of an external molecular force field, (iv) potential numerical instabilities due to reliance on eigendecomposition, and (v) other inefficiencies in the generation of rotamers.

Here, we demonstrate that decoupling the energy and entropy training losses and propagating forces directly from the molecular force field reduces the needed evaluations of the force field by a factor of a thousand; we achieve rare-event sampling with only $10^2$-$10^3$ evaluations of the force field for chicken villin headpiece (PDB ID 1VII), a 35-residue protein domain. We demonstrate a simple method of gradient propagation for an arbitrary external force field, and we implement the AMBER 14 force field (*2*) in pure PyTorch (*16*), as is done in the TorchMD framework (*17*). We include the Generalized Born implicit solvent (*18–22*), which is not present in TorchMD (*17*). We suggest strategies to avoid numerical instabilities and intrinsic biases in the neural network, and we propose a code-efficient method of rotamer sampling that can handle arbitrary molecules while remaining end-to-end differentiable for neural network training. We also present a highly parallel and memory-efficient version of the rotamer sampling algorithm. The result of these improvements is a numerically robust and fast architecture for Boltzmann generators.

**Results**

With our methods, we were able to train neural networks which produced rotameric states with near-native energies, requiring only hundreds to thousands of evaluations of the energy function, rather than the hundreds of thousands to millions in Boltzmann generators, and millions to billions for traditional MD.

Our novel contributions are differentiable rotamer sampling, parallelized differentiable rotamer sampling (Fig. 1, Fig. 2), a thorough guide to adapting OpenMM force fields for PyTorch or TensorFlow use (particularly outlining the calculations necessary for the generalized Born implicit solvent, which is not a feature of the similar framework TorchMD (*17*)), and an *ad hoc* method of using arbitrary force fields for gradients without modification to existing force field code, and without code for custom gradients. We also propose the use of unbiased angle generation methods suitable for neural network training.

*Training a convergent model*
We observed that due to occasional large force gradients, typically due to Lennard-Jones internuclear repulsions, certain training steps would cause rapid divergence of model parameters. Even with regularization techniques including layer normalization, weight decay, force regularization, gradient clipping, and the use of the Adam training algorithm, it was still necessary to adjust the learning rate from the default $\gamma = 10^{-3}$ to $\gamma = 10^{-5}$ or $10^{-6}$. We describe how we performed this tuning in the Methods.

*Initializing Boltzmann generators at the native (or input) state*
We tested the effect of pretraining the neural networks to output $\theta = 0$, or in other words to produce the identity function on the structure. For these experiments (Fig. 3a-b), we trained the neural network for $n_{pre}$ epochs with the loss

$$L_{pre} = \sum_i \theta_i^2 + L_{reg}$$

(1)

where $L_{reg}$ includes fixed terms such as the angle modulus loss in Eq ( 50 ) in the Methods and weight decay regularization.

We observed that initial pre-training was crucial to allow neural networks to converge. With no pre-training epochs, we were unable to train neural networks within 5,000 epochs, and all structures produced appeared highly unphysical, with numerically infinite energies (not shown). However, any number of pre-training epochs above 100 appeared suitable for initialization of the neural networks (Fig. 3a).

*Sampling without entropy*
When we trained neural networks solely on the energy and not the entropy (Fig. 3a-b)

$$L_{energy}(\theta) = U(\theta) + L_{reg}$$

(2)

we observed that while the intrinsic noise in the Adam optimizer allowed for sampling of states which were not global energy minima (Fig. 3a), the resulting neural networks did not reproduce the results of traditional MD at non-zero temperature (Fig. 3b).

*Temperature and entropy, and effect of training length on structure generation*
We observed that by estimating the multivariate circular distribution entropy and training to maximize the entropy and minimize the energy, we were able to reproduce traditional MD protein backbone root mean square fluctuations (Fig. 3c-h), as well as prevent mode collapse (Fig. 4a-c) and sample Boltzmann-distributed states (Fig. 4d-i). For this set of experiments, we used the training loss

$$L_{train}(\theta; T_{NN}) = \begin{cases} \dfrac{U(\theta)}{T_{NN}} - S(\theta) + L_{reg}, & T_{NN} \geq 1 \\ U(\theta) - T_{NN} \cdot S(\theta) + L_{reg}, & T_{NN} < 1 \end{cases} \quad (3)$$

where $T_{NN}$ is the temperature of the system. We chose this form of the loss to maintain the relative contributions and dynamic range of the energy function $U$ and entropy $S$, while preventing exploding gradient contributions from dividing $U$ by a small $T_{NN}$ or multiplying $S$ by a large $T_{NN}$. In the units of our implementation,

$$T_{NN} = R \cdot T \quad (4)$$

with the ideal gas constant $R = 8.314 \times 10^{-3}$ kJ mol$^{-1}$ K$^{-1}$ and physical temperature measured in Kelvins $T$. A human body temperature of 310 K yields a numerical value of $T_{NN} = 2.58$. However, since our estimate of the entropy is only correct asymptotically, the actual numerical values for the temperature may differ in practice. Interestingly, it appears that reproducing the results from traditional MD is a matter of fine-tuning both the length of neural network training and temperature (Fig. 3c, Fig. 4d-i).

*Benchmarking*
We performed basic benchmarking of the rotamer sampler, the parallelized version of the rotamer sampler, and the force field, on the CPU-only of an M1 Max Macbook Pro with 64 GB RAM, and on a NVIDIA Tesla T4 GPU with 16 GB RAM (Fig. 5). These benchmarks were performed on the combined forward and backward passes through the computational graph, to imitate real-world usage. We observed an advantage of up to 10x on the GPU in terms of total throughput of rotamer sampling (Fig. 5a-b). We also compared the original non-parallel dihedral sampler (Algorithm 3) to the parallel dihedral sampling, running on the NVIDIA GPU (Fig. 5c-d), which showed a performance advantage of 4x on the protein we used for these experiments. Finally, we demonstrated that the energy function also demonstrated a performance benefit running on the NVIDIA GPU compared to CPU, with 10x greater performance (Fig. 5e-f).

**Discussion**
By decoupling energy minimization from entropy maximization, we were able to perform Boltzmann sampling with three orders of magnitude fewer calls to the energy function than in the original Boltzmann generators. We were able to reproduce traditional MD results in the form of RMSF from the initial state. Additionally, we provide benchmarks for our algorithms which demonstrate automatic parallelization through PyTorch allowing for significant computational throughput on an NVIDIA GPU.

Unlike in the original Boltzmann generator work, we were able to directly sample internal degrees of freedom (i.e. the dihedral angles), explicitly freezing out all other modes in the molecule. A side effect of our approach was that we did not need to manually remove modes such as overall molecular rotation/translation through eigendecomposition (*14*). Despite removing a large fraction of the degrees of freedom from the molecule, we were still able to reproduce the RMSF profile of the protein computed by traditional MD (Fig. 3c). The methods we present may be useful in examining, for example, protein allostery with far less computation than required by traditional MD.

In terms of performance, it is not surprising that the GPU outperformed the CPU significantly, given a large enough batch size (Fig. 5). We also observed the expected plateau in performance increase due to the saturation of the CUDA driver with simultaneously executing kernels. We discuss the theoretical advantage of the parallelized rotamer sampler in the Methods. Our energy function was not completely optimized, since many of the pairwise computations were computed for both the upper and lower triangles of the distance matrix, resulting in a two-fold redundancy in certain calculations which are symmetric in the particle order. We encountered difficulty with the limited memory on the GPU (the NVIDIA Tesla T4 only has 16 GB of video RAM), particularly with the energy function, whereas the CPU had access to 64 GB of RAM at the cost of compute speed. We did not perform benchmarks on the M1 Max GPU because the PyTorch backend for Apple' Metal Performance Shaders is not complete.

In the original Boltzmann generators proposed by Noé et al., angles were generated as a real number, and invertibility of the networks was ensured through a penalty on angles outside the range $[-\pi, \pi)$ through a squared loss for angles generated outside that range (*14*). This method of angle generation makes it difficult for the network to explore angles near the extremes $-\pi$ and $\pi$, and forces the neural network (a continuous model) to approximate a discontinuous transformation from the angle $\theta$ to the circle $S^1$. In this work, we used the differentiable `atan2` function to generate angles, which does not suffer from these difficulties (Methods). Additionally, the form of the loss biases generation of angles to $\theta = 0$, as we illustrate in a simplified Markov chain model of training (Methods).

One major theoretical limitation of traditional MD which carries over to Boltzmann generators is difficulty in sampling disconnected local energy minima (i.e. metastable states). Fundamentally, the neural networks used in Boltzmann generators are differentiable models which generate $n$ molecular internal coordinates from $m$ latent variables, and are therefore continuous functions from $\mathbb{R}^m$ to $\mathbb{R}^n$ (*23*). Furthermore, the Boltzmann generators originally proposed by Noé et al. (*14*) and in this work generate internal coordinates from the sampling of a single multidimensional Gaussian distribution

$$z \in \mathbb{R}^m \sim N(\mu = 0, \Sigma = I)$$

(5)

centered at $\mu = 0$ with unit standard deviation in each coordinate. In Noé et al., $m$ and $n$ were both set to three times the number of atoms in the protein (i.e. the 3D coordinates). This distribution is spherically symmetric; however in high dimensions, the volume of the unit $m$-ball tends to 0 even for modest values of $m$, implying that the density of the multidimensional Gaussian distribution is highly concentrated near the origin. Meanwhile, the probability density of the molecule's Boltzmann distribution is highly concentrated in disjoint regions of $\mathbb{R}^n$, since the energy minima of a molecule are separated by high energy barriers (low Boltzmann probability). Since $z$ tends to the origin in the latent space, we are asking the neural network to approximate a one-to-many relation, which is not a function, let alone a continuous one. Instead, it may be beneficial to sample $z$ from a sum of Gaussians

$$z \sim \sum_i N(\mu = x_i, \Sigma = I)$$

(6)

and to require that the result of sampling from distinct regions $x_i \neq x_j$ results in internal coordinates $p_i$ and $p_j$ which are also disjoint, such as through a repulsive loss on those pairs of $p_i, p_j$. This method would be analogous to metadynamics sampling (*12*), in which previously generated molecular states are avoided, formulated in the distributional sense. This method is also analogous to k-means clustering, in which each centroid is responsible for representing a single cluster in the data. Alternatively, one could use an ensemble of neural networks, with each neural network responsible for generating Boltzmann-distributed states for a single energy minimum and its neighborhood of conformations.

Another theoretical issue with practical considerations is the size of the neural network necessary to represent the molecule. In the original Boltzmann generator proposed by Noé et al., all $3N$ atomic coordinates are predicted, with principal component analysis to remove the 6 components corresponding to rotations and translations of the entire structure as well as to learn correlations among the degrees of freedom (*14*). Noé et al. also required that the entire neural network be invertible so that exact gradients for their KL divergence between Gaussian priors and Gaussian posteriors may be backpropagated. Thus, the input dimension of their neural networks has the same $3N$ dimensions. Even for small proteins like the chicken villin headpiece we studied, $N = 596$. The neural networks quickly grow in number of trainable parameters. If we restrict the number of hidden units in any layer to a value less than $3N$, then the Jacobian determinant of the neural network transformation will immediately become 0. Therefore, the number of parameters in an invertible network is at least $9N^2$ (3,196,944 for chicken villin headpiece, not including bias parameters). In practice, we require multiple hidden layers with just as many parameters to gain sufficient approximation power for the neural network. Though in principle, the number of trainable may be reduced by choosing fixed values for some of the parameters, such a choice requires further assumptions on the symmetries of the protein.

In this work, we did not assume a Gaussian posterior distribution for the degrees of freedom, and we postulated that many of the degrees of freedom are unimportant to protein dynamics near the energy minima. Though we no longer have a closed form expression for the entropy, we gain an enormous computational advantage in the neural network. Since we used a Gaussian input of dimension 32 (Methods), our networks reproduce a maximum of 32 normal modes in the output. For backbone dihedral sampling, we required only a prediction of the $\phi, \psi, \omega$ angles for each of the 36 residues, leading to a final dimension of 105. Our networks for chicken villin headpiece had 196,692 trainable parameters. It appeared that 32 normal modes were enough to describe the near native structures to high accuracy (Fig. 3c), and we simultaneously accommodated for non-Gaussian posterior distributions for the dihedral angles. In comparison to the original approach proposed by Noé et al., the methods we propose can handle larger proteins in less memory and with fewer trainable parameters. The major caveat is that each protein may require tuning of neural network hyperparameters, especially the minimum dimension of normal modes to predict.

In the case of classical molecular force fields, the high energy barriers tend to be a result of physical singularities in Lennard-Jones and Coulomb potentials at low distance due to internuclear forces, with Lennard-Jones repulsion ($\propto 1/r^{12}$) dominating any electrostatic force ($\propto 1/r^2$) at low distance. One might hope that the high internuclear forces are UV divergences in the classical theory which disappear upon quantization. However, even in the full quantum field theory of the strong nuclear force, while the repulsion we observe is finite, it is still far larger than any of the forces due to electromagnetism, and we must avoid generating states in which the nuclei are too close. In traditional MD and in Monte Carlo methods, umbrella sampling is used to regularize the singularities; we implemented umbrella sampling in this work by directly regularizing the Lennard-Jones forces for nuclear repulsion to a finite, constant repulsive force. In Noé et al., a logarithmic regularization is performed on the total energy. There is additional regularization of the force field when training the neural network, accomplished through gradient clipping (*24*), dropout (*25*), and weight penalties (*16*). Such methods of regularization are forced upon the user in the process of training stable neural networks. In this sense, Boltzmann generators as originally presented are umbrella-sampled.

Despite their shortcomings, Boltzmann generators maintain at least one significant advantage over traditional MD with umbrella sampling. Boltzmann generators implicitly retain some memory of the entire training trajectory, and therefore they may reuse knowledge of the structure between different energy states. For example, an ideal Boltzmann generator learns the correlations among the internal coordinates, correlations which may hold between distinct energy minima. As a result, we may directly estimate the entropy of the internal coordinates, i.e., the elusive conformational entropy.

Training of quantum circuits has recently emerged as a promising technique in molecular biophysics (*26–30*). Example applications of quantum computers include generation of small molecule structures (*27*, *30*), molecular docking (*26*, *28*), and optimization of molecular energies (*29*). Current state-of-the-art quantum computers suffer from noise and low memory capacity. By representing the energy landscape of a molecule in terms of its rotameric degrees of freedom, the differentiable rotamer sampling algorithm significantly reduces the number of parameters necessary to represent the molecule, from the naïve $3N$ coordinates to a much smaller set. Since quantum computers can hold a superposition of states in memory, it may be advantageous to create and train a quantum Boltzmann generator for small molecules such as alanine dipeptide. However, numerous theoretical and practical challenges remain in quantum circuit training, such as the barren plateau problem and the noisiness of current quantum computers (*31*).

Significant challenges remain in the practical use of Boltzmann generators. Even though we were able to reproduce RMSFs with high correlation (>0.7, Fig. 3c), the resulting structures still resemble the native state. We found that the energy landscape of the protein was a highly sensitive function of both the temperature and learning rate; fine-tuning of both appears to be required to produce useful results. However, the methods we present in this work make the fine-tuning process more easily accessible to researchers. Future work may also examine the effect of neural network architecture and other hyperparameters on angle generation since we did not study that effect here. Finally, it may be possible to further accelerate certain computations with a field-programmable gate array, which can be configured on a per-protein basis to perform rotamer sampling and energy computation.

In conclusion, we present a comprehensive toolkit of differentiable methods for molecular science. Our contributions include: *ad hoc* propagation of forces from an arbitrary force field for cases in which rewriting the force field is infeasible, differentiable and parallel rotamer sampling/protein fragment assembly, a guide to writing molecular force fields in a differentiable programming framework, decoupling of energy and entropic estimation, and mathematical results on 3D point cloud alignment and 3D rotation representation which can be applied to problems in molecular geometry. We additionally address potential sources of bias in molecular structure generation and outline the approach to remaining sources of bias which we did not implement. We demonstrate that our methods are efficiently implementable on CPU and GPU, and mathematically sound. We hope that other researchers will find these methods and the accompanying reference code useful in investigating molecular energy landscapes.

**Acknowledgements:**

**Funding:** We acknowledge support from the National Institutes of Health (NIH) 1R35 GM134864, 1RF1 AG071675, 1R01 AT012053, the National Science Foundation 2210963, and the Passan Foundation.

**Author contributions:**
C.M.S: conceptualization, data curation, formal analysis, investigation, methodology, software, validation, visualization, writing – original draft, writing – review & editing.
J.W.: investigation, methodology, software, validation, writing – review & editing.
N.V.D.: formal analysis, funding acquisition, investigation, methodology, project administration, resources, supervision, validation, visualization, writing – review & editing.

**Competing interests:**
We declare that we have no competing interests which might affect the results or conclusions presented in this work.

**Data and materials availability:**
All code (Jupyter notebooks, Python scripts) to reproduce the results in this work are included in the Supplemental Materials (diffrot-code.zip), and we will maintain a BitBucket repository to address errors and bugs (https://bitbucket.org/dokhlab/diffrot-manuscript/). Non-essential pretrained model weights are available upon request to the corresponding author (dokh@psu.edu). We also include example movies generated by our rotameric Boltzmann generators (Movies 1 and 2).

**Supplementary materials**
Materials and methods
Supplementary Table S1
Code (diffrot-code.zip)
Movie 1 and 2 (*.mp4)
Algorithms 1-6

**Figures**

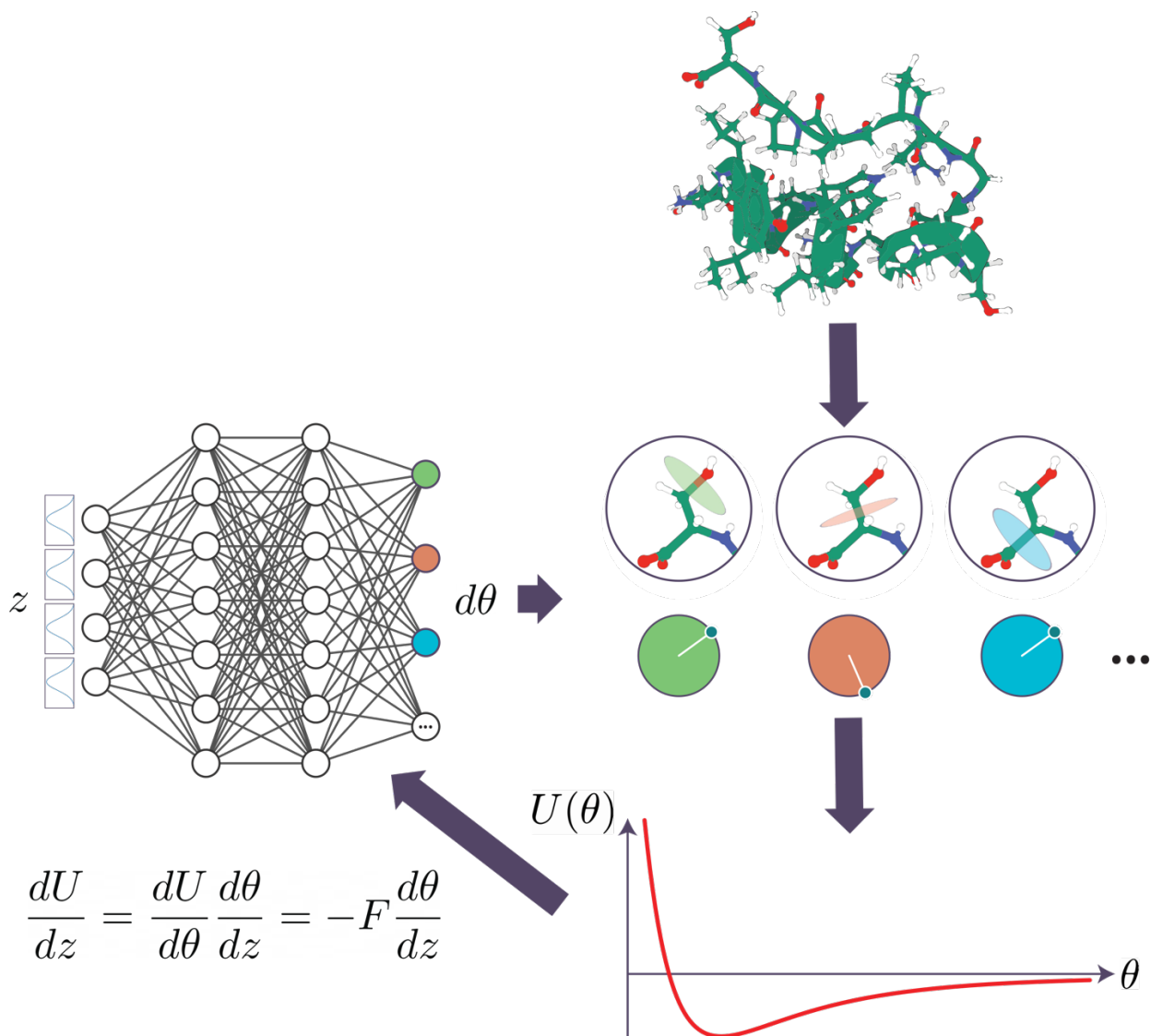

Fig. 1: Differentiable rotamer sampling for Boltzmann generators. Given an arbitrary macromolecule, we identify all the rotameric degrees of freedom; in the case of a protein these degrees of freedom are the dihedral angles of backbones and side chains). We use a neural network to generate changes to the dihedral angles, and perform the rotations on the protein structure. Finally, we evaluate the energy and forces of the resulting structure with respect to a molecular force field, and backpropagate the gradients through the sampling process to the neural network. In this way, we can train the neural network to produce states with low energy, allowing for the study of potential stable conformations of the macromolecule.

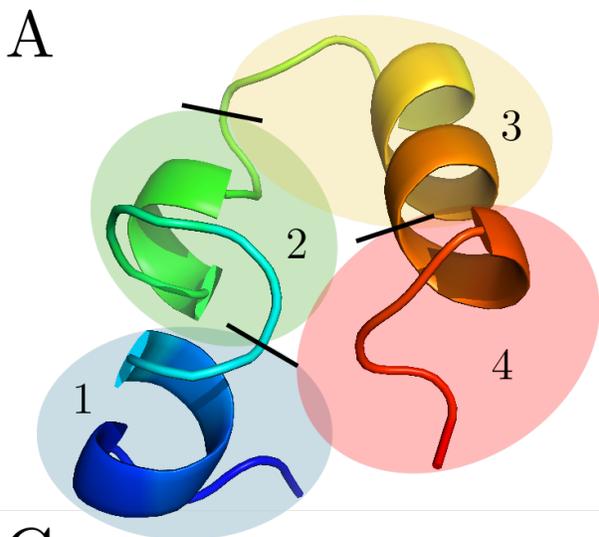
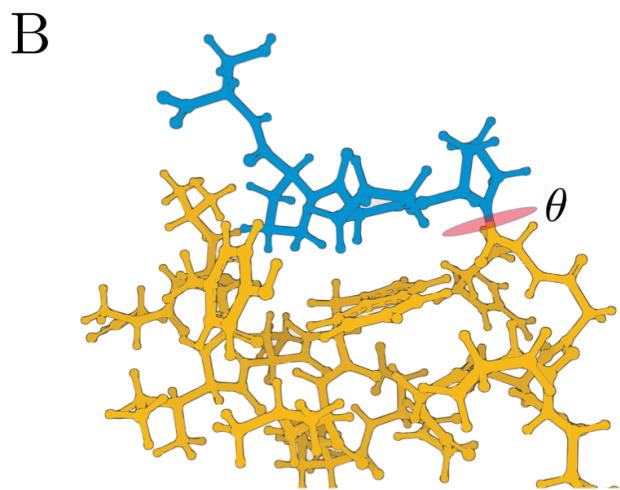
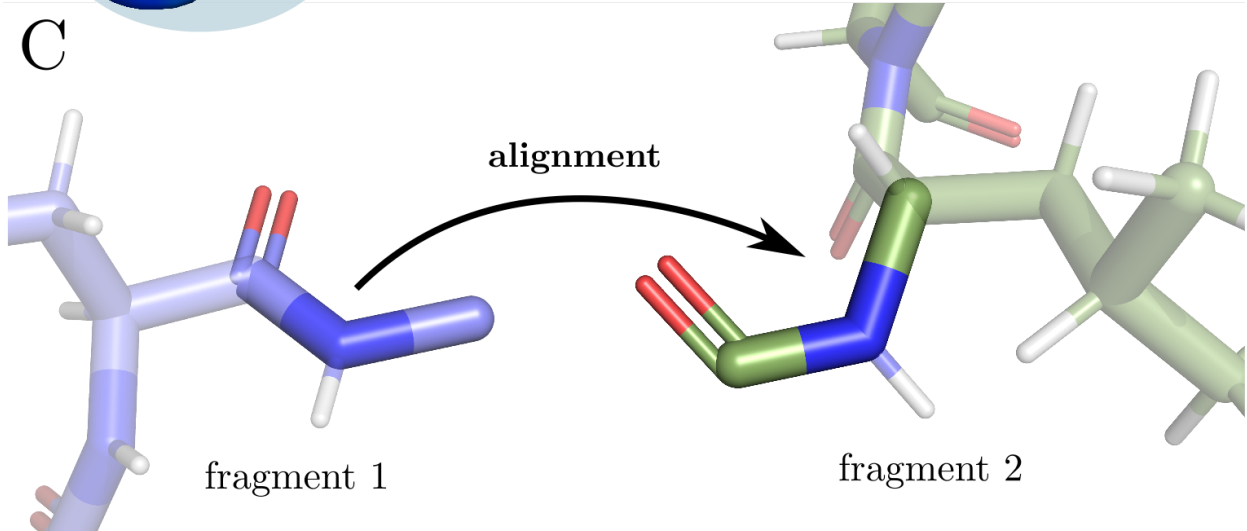
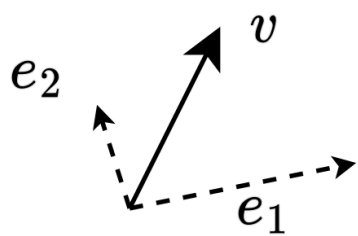
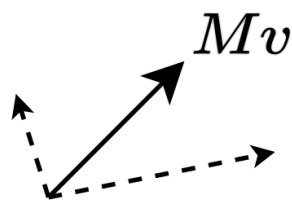
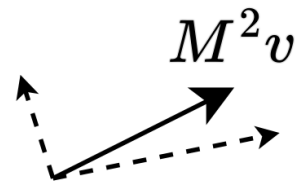
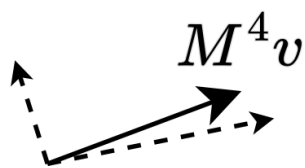
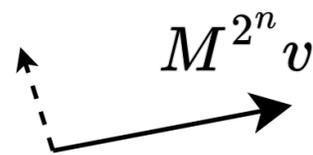

Fig. 2: **Differentiable rotamer sampling workflow.** (a) The protein (chicken villin headpiece, PDB ID 1VII) is split into fragments, whose rotamers can be sampled in parallel. (b) For each rotameric degree of freedom, we split the macromolecule into two connected components at the associated bond, using a depth-first search on the graph of bonds. We then rotate (red) the smaller connected component (blue) about the bond axis by the neural network output $\theta$, using Rodrigues' formula. (c) To combine fragments after dihedrals have been sampled for each fragment, we use an alignment algorithm (Kabsch or quaternionic) on specific atoms in the backbone. Through this method, we can therefore explore the energy landscape of the macromolecule solely as a function of its internal, rotameric degrees of freedom. Because each step of this method is differentiable, we can backpropagate gradients through the rotamer sampling process to provide derivatives for $\theta$. (d) Intuition for exponential acceleration of power iteration. We used repeated matrix squaring to achieve high matrix powers, leading to exponential acceleration of the traditional power iteration algorithm for finding the largest magnitude eigenvalue and associated eigenvector. With higher powers of $M$ multiplying an arbitrary vector $v$, the $e_2$ component is gradually suppressed.

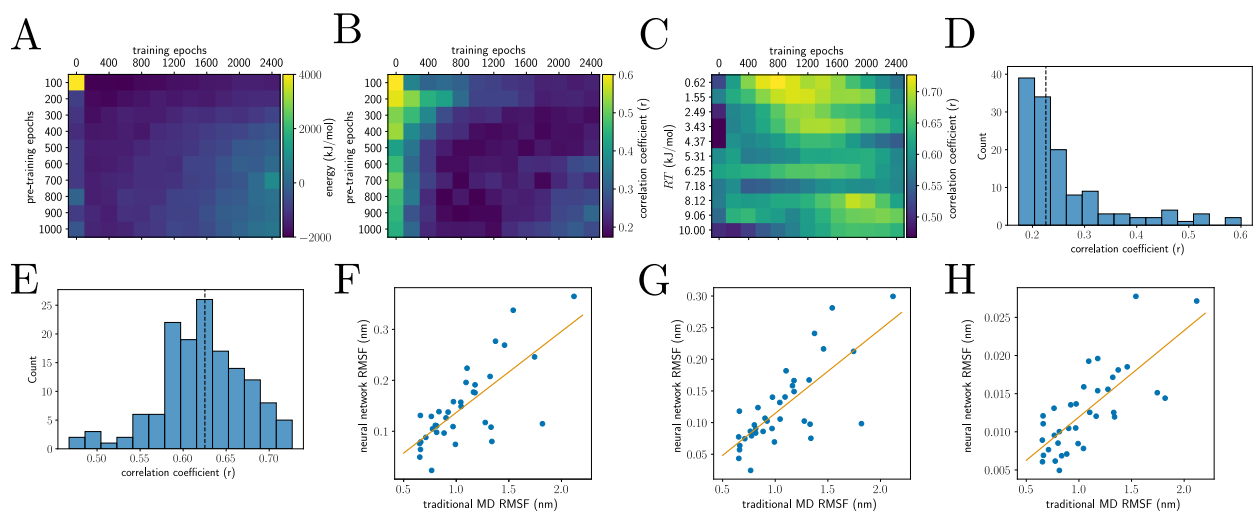

Fig. 3: **Evaluating the consequences of neural network pre-training and entropy (chicken villin headpiece, PDB ID 1VII).** We only sampled the protein backbone dihedrals. (a)-(b) Examining the effect of pre-training to produce the native structure, as well as the role of entropy in neural network training. (a) The minimum energy structure produced from ~600 structures across 3 trained networks; not trained on the entropy. (b)-(c) Correlation coefficients between traditional MD alpha-carbon root mean square fluctuations (RMSF) and neural network RMSF. (b) neural networks trained without entropic estimation; (c) neural networks trained on the full loss described in Results. Entropy training is necessary to reproduce traditional RMSFs. (d) Distribution of all correlation coefficients from (b), with median of 0.23 (dashed line). (e) Distribution of all correlation coefficients from (c), with median of 0.63 (dashed line). We performed a Fisher r-to-z transform on the data for (d) and (e) and performed a t-test for unequal variances, ($t = -39.78$, $p < 0.001$). Training with entropy (e) better reproduces relative RMSFs than without (d). (f)-(h) Traditional MD RMSF versus the neural network RMSF for the four networks with the highest correlation. (f) RT=0.62, 1200 epochs, r=0.73; (g) RT=0.62, 1000 epochs, r=0.75; (h) RT=9.06, 200 epochs, r=0.75.

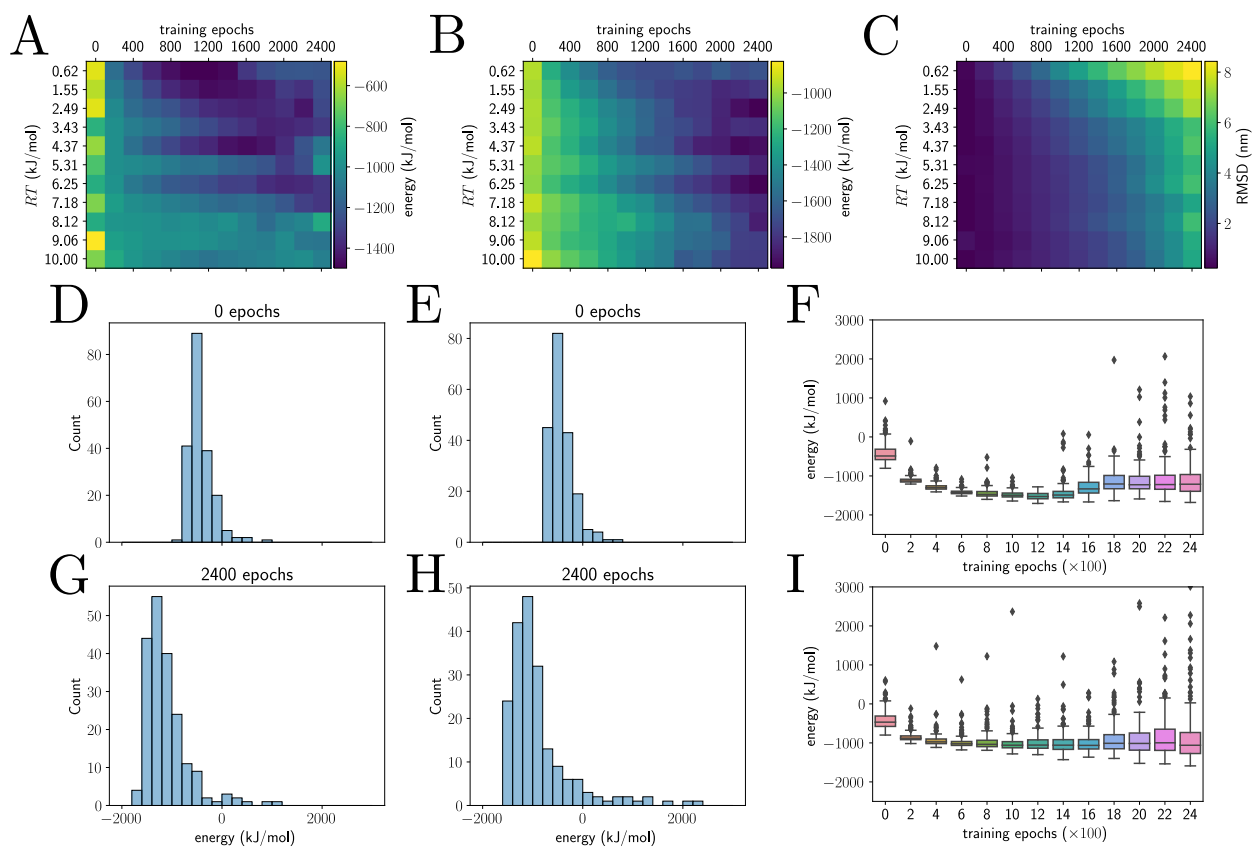

Fig. 4: **Training characteristics across a range of temperatures.** In these plots, we examine the effect of training on the energies of output structures; for rotamer sampling, we only sampled the protein backbone dihedrals. (a)-(c), for each value of the temperature, we trained three neural networks, first on a loss to reproduce the native structure for 1,000 epochs, and then by the Boltzmann loss of the full energy and entropy for up to 2,400 epochs. Colorbars indicate the numerical value for each square. (a) Each square represents the median energy of ~600 generated structures across all three networks. (b) Minimum energy of the structures. (c) The median RMSD from the native structure. (d) and (e) Initial energy distribution at $RT = 0.62$ kJ/mol and $10$ kJ/mol respectively. (g) and (h) Final energy distribution at the same temperatures. (f) Energy distributions as a function of training epochs for one of the networks we trained at $RT = 0.62$ kJ/mol. (i) Energy distributions as a function of training epochs for one of the networks we trained at $RT = 10$ kJ/mol. Higher temperatures homogenize the sampled structures in terms of energy, in (a)-(b). In (d)-(i), we observe that training the neural networks equilibrated within a few hundred epochs of training.

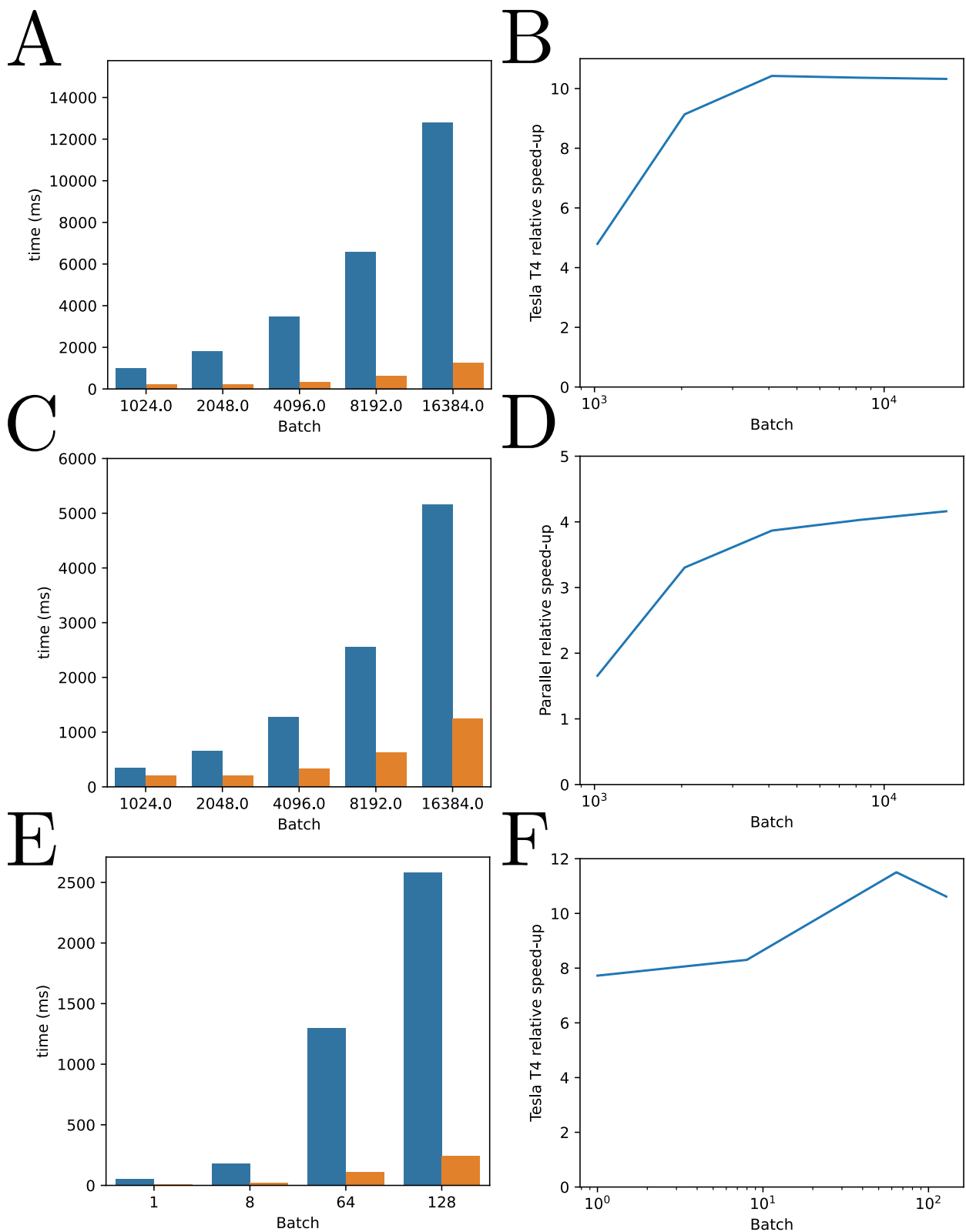

Fig. 5: **Benchmarking of rotamer sampler and the energy function written in PyTorch.** (a) and (b) Comparing performance of the parallel dihedral sampler on chicken

villin headpiece (PDB ID 1VII), split into four fragments. We compared CPU-only computation on an M1 Max MacBook Pro (blue) to a NVIDIA Tesla T4 GPU with 16 GB RAM (orange). We observed a plateau above a batch size of 4096 in performance increase, indicating saturation of the GPU with CUDA kernels. (c) and (d) Comparing the performance of the non-parallel (blue) and parallel (orange) versions of the dihedral sampler. (e) and (f) Comparing the M1 Max (blue) versus the NVIDIA GPU (orange) on the energy function. For a batch size of 256 and higher, the 16 GB GPU ran out of memory.

**Materials and methods**

*Ad hoc propagation of forces*

Since MD force fields are optimized and parallelized to calculate *forces* in addition to energies, we propose using the forces directly as gradients for neural network training. Assume that a neural network $N$ produces atomic positions $x$ from Gaussian random variables $z$ and neural network parameters $p$:

$$x = N(z; p) \tag{7}$$

We sought to optimize $p$ with respect to the energy function

$$U_{MD}(x) = U(N(z; p)) \tag{8}$$

In other words, we sought to calculate the gradients

$$\frac{\partial U}{\partial p_i} = \sum_j \frac{\partial U}{\partial x_j} \cdot \frac{\partial x_j}{\partial p_i} = \sum_j \frac{\partial U}{\partial x_j} \cdot \frac{\partial (N(z;p))_j}{\partial p_i} \tag{9}$$

However, by definition, the forces are the gradients of the energy with respect to the position

$$F_j = -\frac{\partial U_{MD}}{\partial x_j} \tag{10}$$

Therefore, we do not need to backpropagate gradients through the external molecular force field to provide gradients with respect to $p$. Instead of differentiating $U_{MD}$ to train the network, we propose to use the loss

$$U_{new} = -\sum_j F_j \cdot x_j = -\sum_j F_j \cdot (N(z;p))_j \tag{11}$$

Notice that $F_j$ is no longer a function of $x$ (and therefore not of $z$ or $p$ either), since we are using it as a constant input from the molecular force field. Taking a derivative,

$$\frac{\partial U_{new}}{\partial x_i} = -\sum_j F_j \cdot \frac{\partial x_j}{\partial x_i} = -F_i = \frac{\partial U_{MD}}{\partial x_i} \tag{12}$$

Since the gradients for $U_{new}$ are the same as for $U_{MD}$, there is no theoretical difference between training on $U_{new}$ instead of on $U_{MD}$. There is no need for a statistical estimate

(i.e. KL divergence) of the gradients with respect to the force field energy, as was done in the original work by Noé et al (*14*).

This method is shown in Algorithm 1: FFDiff.

*Physical interpretation of $U_{new}$*
We can show that $U_{new}$ has a physical meaning beyond being a convenient loss function.

First, we note that $U_{new}$ is manifestly rotation-invariant. It is the sum of the inner products of vectors in Euclidean space; rotation of the coordinate system rotates both the position and force contravariantly, leaving the inner product unchanged.

Second, $U_{new}$ is translation invariant. Say that we translate a system of particles by a vector $p$. Then in the new primed coordinates, $U'_{new}$ is

$$U'_{new} = -\sum_j F'_j \cdot x'_j = -\sum_j F_j \cdot (x_j - p)$$

(13)

assuming the intramolecular forces are unchanged ($F' = F$) with translation ($x' = x - p$).

Splitting up the net forces $F_j$ into all pairwise forces $F_{ji}$ (the force exerted on particle $j$ by particle $i$), and then splitting up the sum, we may write

$$U'_{new} = -\sum_{i>j} F_{ji} \cdot (x_j - p) - \sum_{j>i} F_{ji} \cdot (x_j - p) - \sum_{i=j} F_{ji} \cdot (x_j - p)$$

(14)

By Newton's third law (assuming the molecular force field is conservative), every force must be paired with an equal and opposite force, so $F_{ji} = -F_{ij}$, and therefore the third sum is 0 ($i = j$). The remaining sums are over the upper and lower triangles of $F_{ji}$.

$$U'_{new} = -\sum_{i>j} F_{ji} \cdot (x_j - p) - \sum_{j>i} F_{ji} \cdot (x_j - p)$$

(15)

By linearity,

$$U'_{new} = -\sum_{i>j}[F_{ji} \cdot x_j - F_{ji} \cdot \boldsymbol{p}] - \sum_{j>i}[F_{ji} \cdot x_j - F_{ji} \cdot \boldsymbol{p}]$$

$$= -\sum_{i>j} F_{ji} \cdot x_j - \sum_{j>i} F_{ji} \cdot x_j - \boldsymbol{p} \cdot \sum_{i>j} F_{ji} - \boldsymbol{p} \cdot \sum_{j>i} F_{ji}$$

$$= -\sum_{i>j} F_{ji} \cdot x_j - \sum_{j>i} F_{ji} \cdot x_j - \boldsymbol{p} \cdot \left( \sum_{i>j} (F_{ji} + F_{ij}) \right)$$

(16)

Finally, we apply Newton's third law again so that the third term cancels, and therefore $U'_{new}$ has no dependence on $\boldsymbol{p}$ and therefore it is translation-invariant.

To interpret the physical meaning of $U_{new}$, it suffices to examine a simple system. Say there is a pair of particles connected by a spring, with particle 1 at the origin ($x_1 = [0, 0, 0]$) and particle 2 at $x_2 = [1, 0, 0]$. Assume that the spring is compressed, so that the force pushes the particles away from each other: $F_1 = [-k, 0, 0]$ and $F_2 = [k, 0, 0]$. In this case, we can directly calculate

$$U_{new} = -F_1 \cdot x_1 - F_2 \cdot x_2 = -[-k, 0, 0] \cdot [0, 0, 0] - [k, 0, 0] \cdot [1, 0, 0] = -k$$

(17)

Hence, if the spring is instead stretched beyond its equilibrium length, then $U_{new}$ is positive. Since $U_{new}$ is invariant under rotation and translation of the system, our choice of coordinate system was irrelevant. Generalizing to many pairs of particles and many springs, the interpretation for $U_{new}$ is that it measures the total compressive energy of the system, with negative values indicating the system is compressed with respect to equilibrium and positive values indicating the system is stretched with respect to equilibrium.

*Differentiable force fields*
Using OpenMM 7.7 (*32*) as our reference, we rewrote force field terms from OpenMM in terms of pure PyTorch operations, allowing for automatic differentiation of the molecular energy without the memory overhead of repeatedly transferring positional data between PyTorch (*16*) and OpenMM. Our implementation creates a custom PyTorch function for a provided molecule which stores the computational graph necessary to reproduce its energy. After it is prepared, the PyTorch energy function requires the user only to supply the coordinates (in nanometers) of each atom, in the same order as presented in the input molecular file. From the user-supplied atomic positions $x_i$, we calculate all pairwise displacements

$$s_{ij} = x_i - x_j$$

(18)

and Euclidean distances

$$d_{ij} = |s_{ij}| = |x_i - x_j| = \sqrt{\sum_k \left((x_i)_k - (x_j)_k\right)^2}$$

(19)

For the all-atom AMBER 14 force field (2), we implemented: (1) harmonic bond lengths, (2) harmonic bond angles, (3) pairwise atomic Coulomb interactions, (4) pairwise atomic Lennard-Jones potentials, (5) periodic backbone dihedral (torsional) angle energies, and (6) generalized Born implicit solvent energies.

First, we imported the molecular topology into OpenMM, which assigns parameter values for each atom or tuple of atoms. Second, we referenced the OpenMM documentation to implement each of the forces.

Harmonic bond lengths: the AMBER force field provides particle indices representing a covalent bond $b = (i, j)$, the bond length $l_b$, and the bond strength $k_b$. For each bond, we calculate an energy term with these parameters and the pairwise atomic distances:

$$E_{harmonic\ bonds} = \sum_b \frac{1}{2} k_b (d_{ij} - l_b)^2$$

(20)

In OpenMM, the parameters are contained in a HarmonicAngleForce object.

Harmonic bond angles: for every triple of atoms which are covalently connected in a linear chain, the AMBER force field provides particles indices $t = (i, j, k) \in T$ in which $j$ is the middle atom, a bond angle $\phi_t$, and a bond angle strength $k_t$. We calculate the angle $\theta_{ijk}$ formed between $s_{ji}$ and $s_{jk}$ (the order of the indices is important) through the dot product identity

$$\cos \theta_{ijk} = \frac{s_{ji} \cdot s_{jk}}{|s_{ji}||s_{jk}|} = \frac{s_{ji} \cdot s_{jk}}{d_{ji} d_{jk}}$$

(21)

The corresponding energy term is then:

$$E_{harmonic\ angles} = \sum_t \frac{1}{2} k_t (\theta_{ijk} - \phi_t)^2$$

(22)

In OpenMM, the parameters are contained in a HarmonicBondForce object.

Pairwise atomic Coulomb and Lennard-Jones interactions: for every pair of atoms, there is an electrostatic force as a function of the interatom distance $d_{ij}$ due to estimated partial charges $q_i$ from AMBER.

$$E_{electrostatic} = \sum_{i<j} \frac{q_i q_j}{4\pi\epsilon_0 d_{ij}}$$

(23)

For every pair of atoms, there is a Lennard-Jones potential which approximates long-range dipole-dipole attractions and short-range nuclear repulsions. AMBER provides an energy scale $\epsilon_{ij}$ and interaction distance $\sigma_{ij}$ for each pair of particles.

$$E_{LJ} = \sum_{i<j} \frac{\epsilon_{ij}}{4} \left[ \left(\frac{\sigma_{ij}}{d_{ij}}\right)^{12} - \left(\frac{\sigma_{ij}}{d_{ij}}\right)^{6} \right]$$

(24)

Certain Lennard-Jones and Coulomb interactions are often ignored or modified for atoms which are within a certain number of bonds of each other. These modifications can be thought of as setting elements of the matrices $\epsilon_{ij}$, $\sigma_{ij}$, and $d_{ij}$ to specific values and are included in the OpenMM implementation of AMBER.

During regularization of the force field, we adjusted the Lennard-Jones potential so that for all particles below a certain distance, the force was a constant repulsive one. Given a maximum force magnitude for internuclear repulsion $F_m$, and for each Lennard-Jones interaction, we first computed the magnitude of the force for that interaction

$$F_{LJ,ij}(d_{ij}) = \frac{\epsilon_{ij}}{4 d_{ij}} \left[ 6 \left(\frac{\sigma_{ij}}{d_{ij}}\right)^{6} - 12 \left(\frac{\sigma_{ij}}{d_{ij}}\right)^{12} \right]$$

(25)

For numerical stability, we computed $\left(\frac{\sigma_{ij}}{d_{ij}}\right)^{6}/d_{ij}$ and $\left(\frac{\sigma_{ij}}{d_{ij}}\right)^{12}/d_{ij}$ in the given order of operations rather than taking the ratio of the powers $\frac{\sigma_{ij}^6}{d_{ij}^7}$ and $\frac{\sigma_{ij}^{12}}{d_{ij}^{13}}$. At $r_{min,ij} = \sqrt[6]{2}\sigma$, the Lennard-Jones interaction achieves its minimum value, and thus $F_{LJ,ij}(r_{min,ij}) = 0$. This is the only minimum of the Lennard-Jones potential, and it can be seen that $F_{LJ,ij}$ is a monotonically increasing function of the distance, with an asymptote at $d_{ij} = 0$ which tends to negative infinity. Therefore to find the value $r_{max\ grad,ij}$ at which the magnitude of the repulsive force is equal to $F_m$, we used the bisection method for root finding (*33*). We did not use the Newton-Raphson method or other methods which require the second derivative of the Lennard-Jones potential due to numerical stability.

For a given value of $0 \leq F_m < \infty$, we know that the solution of the equation

$$\left|F_{LJ,ij}(d_{ij})\right| - F_m = 0$$

lies in the interval $(0, r_{min,ij}]$. Therefore by starting with the initial guess $d_{ij}^0 = r_{min,ij}/2$, we may bisect the interval repeatedly to determine the approximate value of the zero ,

depending on if $|F_{LJ,ij}(d_{ij})| > F_m$ or $|F_{LJ,ij}(d_{ij})| < F_m$. We performed this bisection $n_{bisect} = 100$ times, which should ensure a negligibly small error for $r_{max\,grad,ij}$. Since $\sigma_{ij}$ was measured on the nanometer scale and numerical values for $\sigma_{ij}$, the bisection method should be accurate to roughly 1 part in $2^{n_{bisect}}$, or $10^{-30}$ nanometers.

We defined the regularized energy as

$$E_{LJ,reg} = \sum_{i<j} \begin{cases} \frac{\epsilon_{ij}}{4}\left[\left(\frac{\sigma_{ij}}{d_{ij}}\right)^{12} - \left(\frac{\sigma_{ij}}{d_{ij}}\right)^{6}\right], & \text{if } d_{ij} > r_{max\,grad,ij} \\ E_{0,ij} - F_m d_{ij}, & \text{otherwise} \end{cases}$$

(26)

Where $E_{0,ij}$ was chosen so that the $E_{LJ,reg}$ is continuous at $d_{ij} = r_{max\,grad,ij}$. By inspection, the derivative of the regularized potential at close internuclear distances is $-F_m$, as desired.

In OpenMM, parameters for Coulomb and Lennard-Jones interactions, as well as any modifications, are contained in a NonbondedForce object.

Periodic backbone dihedral (torsional) angle energies: for every linear chain of four atoms $c = (i,j,k,l)$, AMBER provides a periodic force for the dihedral angle, which is defined as the angle that $s_{ji}$ makes with respect to $s_{kl}$ when viewed in the plane whose normal is $s_{jk}$, defined by a periodicity $n_c$, energy scale $\epsilon_c$, and phase shift $\phi_c$.

The dihedral angle $\theta_c$ is calculated by first normalizing $s_{jk}$:

$$\hat{s}_{jk} = \frac{s_{jk}}{d_{jk}}$$

(27)

and projecting $s_{ji}$ and $s_{kl}$ onto the plane of interest:

$$\tilde{s}_{ji} = (1 - s_{ji} \cdot \hat{s}_{jk})s_{ji}$$
$$\tilde{s}_{kl} = (1 - s_{kl} \cdot \hat{s}_{jk})s_{kl}$$

(28)

Then the dot product identity states that

$$\tilde{s}_{ji} \cdot \tilde{s}_{kl} = |\tilde{s}_{ji}||\tilde{s}_{kl}| \cos\theta_c$$

(29)

To avoid numerical instabilities and to recover the dihedral angle from the full range $[-\pi, \pi)$ (34), we compute

$$(\hat{s}_{jk} \times \tilde{s}_{ji}) \cdot \tilde{s}_{kl} = |\tilde{s}_{ji}||\tilde{s}_{kl}| \sin \theta_c$$

(30)

which is a special case of the polar sine, and finally

$$\theta_c = \text{atan2} \frac{\tilde{s}_{ji} \cdot \tilde{s}_{kl}}{(\hat{s}_{jk} \times \tilde{s}_{ji}) \cdot \tilde{s}_{kl}}$$

(31)

The torsional energy is then:

$$E_{torsion} = \sum_c \epsilon_c [1 + \cos(n_c \theta_c - \phi_c)]$$

(32)

In OpenMM, the parameters are contained in a PeriodicTorsionForce object.

<u>Generalized Born implicit solvent</u>: we used the generalized Born implicit solvent with improved parameters (*22*, *35*), labeled GBn2 in OpenMM or igb=8 in AMBER. The functional form for this energy is complicated, so we only present the calculations necessary to reproduce the energy. Global parameters for the implicit solvent are cutoff$_{neck}$ = 0.68, scale$_{neck}$ = 0.826836, offset = 0.0195141, $\epsilon_{solute}$ = 1, $\epsilon_{solvent}$ = 78.5, and integral corrections $\eta$ = 28.3919551 and $\theta$ = 0.14; per particle and per pair parameters are Born radii radius$_i$, Born/van der Waals cutoff adjustments or$_i$ and sr$_i$, partial charges $q_i$, effective radii scaling parameters ($\alpha_i, \beta_i, \gamma_i$), and pairwise neck integral parameters $d_{0,ij}$ and $m_{0,ij}$. Given pairwise particle distances $d_{ij}$, we calculate

$$D_{ij} = d_{ij} - sr_j$$

(33)

$$L_{ij} = \max(or_i, D_{ij})$$

(34)

$$U_{ij} = d_{ij} + sr_j$$

(35)

$$I_{vdw,ij} = \begin{cases} \frac{1}{2}\left[\frac{1}{L_{ij}} - \frac{1}{U_{ij}} + \frac{1}{4}\left(d_{ij} - \frac{sr_j^2}{d_{ij}}\right)\left(\frac{1}{U_{ij}^2} - \frac{1}{L_{ij}^2}\right) + \frac{1}{2d_{ij}}\log\frac{L_{ij}}{U_{ij}}\right], & d_{ij} > sr_i - or_i \\ 0, & \text{otherwise} \end{cases}$$

(36)

$$I_{neck,ij} = \begin{cases} \frac{m_{0,ij}}{1 + 100(d_{ij} - d_{0,ij})^2 + 300{,}000(d_{ij} - d_{0,ij})^6}, & d_{ij} < \text{radius}_i + \text{radius}_j + \text{cutoff}_{neck} \\ 0, & \text{otherwise} \end{cases}$$

(37)

$$I_{ij} = I_{vdw,ij} + \text{scale}_{neck}I_{neck,ij}$$

(38)

For each particle, note that $I_{ii} = 0$. We then compute

$$I_i = \sum_j I_{ij}$$

(39)

$$\psi_i = I_i \text{or}_i$$

(40)

$$B_i = \left[\frac{1}{\text{or}_i} - \frac{\tanh(\alpha_i\psi_i - \beta_i\psi_i^2 + \gamma_i\psi_i^3)}{\text{radius}_i}\right]^{-1}$$

(41)

$$\text{or}_{\text{offset},i} = \text{or}_i + \text{offset}$$

(42)

$$f_{GB,ij} = \sqrt{d_{ij}^2 + B_iB_j \exp\left(-\frac{d_{ij}^2}{4B_iB_j}\right)}$$

(43)

The implicit solvent energy is then

$$E_{GB} = -\frac{1}{2}\sum_i \frac{q_i^2}{4\pi\epsilon_0 B_i}\left[\frac{1}{\epsilon_{solute}} - \frac{1}{\epsilon_{solvent}}\right] + \sum_i \eta\left(\text{or}_{\text{offset},i} + \theta\right)^2 \left(\frac{\text{or}_{\text{offset},i}}{B}\right)^6$$
$$- \sum_{i<j} \frac{q_1 q_2}{4\pi\epsilon_0 f_{GB,ij}}\left[\frac{1}{\epsilon_{solute}} - \frac{1}{\epsilon_{solvent}}\right]$$

(44)

In OpenMM, the parameters are contained in a CustomGBForce object. The implicit solvent modifies the Coulomb energies computed earlier.

We computed all non-indexed sums on pairs of particles using the full matrix format, which duplicated some of the necessary computations. We verified that energies and gradients from our PyTorch implementation matched that of OpenMM to within 5% (see Code).

*Avoiding singularities in the energy function during backpropagation*
Numerical singularities in gradients arose due to true singularities in the energy function and artificial singularities due to the computational graph.

True singularities in the gradient arose during both the forward and backward passes due to square roots, logarithms, and negative powers. In all cases in our energy function, we are guaranteed that the argument of the function is nonnegative. Therefore, we avoided true singularities by adding a small positive offset $\epsilon$ to all affected computations.

For computations which were performed but later discarded, such as those involving the diagonal of the distance matrix, we manually set specific terms to fixed constant values using PyTorch's masked fill function so that during backpropagation, the gradient for those terms would be fixed to 0 instead of NaN.

*Preparation of macromolecule graph metadata for rotameric sampling*
Given an arbitrary macromolecule whose atoms are covalently bonded into a single connected structure, we sought to modify only the dihedral angles.

First, we created an undirected graph (*36*) $G = (V, E)$ of the macromolecule, with atoms as the nodes $V$ and covalent bonds as the edges $E \subseteq V \times V$.

Second, many macromolecules contain cycles which reduce the number of degrees of freedom by one, so we performed a depth-first search (which generates a tree) to break cycles at a single bond and retaining all other bonds as dihedral degrees of freedom: $T = (V, E_T)$.

Third, we removed all leaf nodes and edges terminating on leaves, since the bonds corresponding to such edges do not represent a dihedral angle: $T_{dih} = (V_{dih}, E_{dih})$. For each remaining edge $e \in E_{dih}$, we assigned an output of the neural network to control the dihedral angle for that bond.

Finally, for each dihedral edge $e \in E_{dih}$, we used depth-first search to calculate the two connected components of $T_{dih}$ that result with the removal of $e$. We recorded the smaller of the two connected components, $CC(e) \subseteq V$, as the atoms which we would rotate about the dihedral axis represented by the $e$.

This method is shown in Algorithm 2: PrepRot.

*Differentiable rotamer sampling*
We started with the $N \times 3$ matrix of positions $x_{ij}^{(0)}$ ($1 \leq i \leq N, 1 \leq j \leq 3$) of all the atoms in the macromolecule. For each of the angles $\theta_m$ predicted by the neural network, we selected the corresponding edge $e_m = (u_m, v_m)$, where $u_m$ and $v_m$ represent the two atoms forming the covalent bond. We calculated the axis of rotation with components

$$k_m = p_{u_m} - p_{v_m}$$

(45)

using the positions of atoms $u_m$ and $v_m$, which have components $(p_{u_m})_j = x_{u_m j}^{(m)}$ and $(p_{v_m})_j = x_{v_m j}^{(m)}$. We did not prefer a specific orientation for each rotation axis, since we predicted the full $2\pi$ range for each $\theta_m$. We also computed the centroid of the bond as the origin for rotation:

$$c_m = \frac{1}{2}(p_{u_m} + p_{v_m})$$

(46)

We normalized $k_m$ to a unit vector, $\hat{k}_m$. We used Rodrigues' rotation formula(37) for rotation around an axis by an angle to calculate a rotation matrix $R(\theta_m, \hat{k}_m)$:

$$R(\theta_m, \hat{k}_m) = I + (\sin \theta_m)K_m + (1 - \cos \theta_m)K_m^2$$

(47)

$$K_m = \begin{bmatrix} 0 & -k_z & k_y \\ k_z & 0 & -k_x \\ -k_y & k_x & 0 \end{bmatrix}$$

(48)

We then took the previously calculated connected component $CC(e_m)$, translated all the particles in that connected component by $-c_m$, applied the rotation matrix $R_m$, and translated the selected particles back by $c_m$:

$$x_{ij}^{(m+1)} = \begin{cases} ([R_m \cdot (p_i - c_m)] + c_m)_j, & \text{if } i \in CC(e_m) \\ x_{ij}^m, & \text{otherwise} \end{cases}$$

(49)

After performing this transformation for the $M$ dihedrals we wish to sample, we return our final position matrix as the output, $x_{ij}^{(M+1)}$.

This method of differentiable rotamer sampling is shown in Algorithm 3: DiffRot.

*Bias-free, continuous representation of dihedral angles*
We used simple feedforward neural networks, with a continuous representation of angles for the output (23). To avoid biasing predicted angles, we did not predict angles directly. For each dihedral angle $\theta_m$, we used our neural network to predict two parameters, $(x_m, y_m)$, and calculated $\theta_m = \text{atan2}(y_m, x_m)$. atan2 and its derivative are well-defined for all $(x_m, y_m)$ in all four quadrants and produces an angle in the range $[-\pi, \pi)$. To regularize our neural networks and prevent $(x_m, y_m)$ from drifting to the origin or infinity, we added a loss to our training cost, with weight $\epsilon_{xy}$:

$$L_{xy} = \epsilon_{xy} \sum_m (x_m^2 + y_m^2 - 1)$$

(50)

This training cost has the advantage of being rotationally symmetric so that there is no preferred angle. For comparison, the regularization loss

$$L_\theta = \epsilon_\theta \left( \sum_{\theta_m > \theta_{max}} \theta_m^2 + \sum_{\theta_m < \theta_{min}} \theta_m^2 \right)$$

(51)

will bias every angle $\theta_m$ toward $\frac{1}{2}(\theta_{max} + \theta_{min})$, since the network is penalized for exploring the space near $\theta_{max}$ and $\theta_{min}$ *(14)*.

To show mathematically that the neural network training is biased in the latter case, even for values of $\theta$ within the range $[\theta_{max}, \theta_{min}]$, we model the training process as a Markov chain *(38)*, and we also assume that $\theta$ can only take on a discrete set of values in the range $[\theta_{max}, \theta_{min}]$ of size $N$, or the values $\theta_{min} - \epsilon$ or $\theta_{max} + \epsilon$. As a toy model, we assume that the training algorithm makes a transition $\theta \to \theta - \epsilon$ or $\theta \to \theta + \epsilon$ with equal probability while $\theta$ is within the range, and by the probability 1 transitions

$$\theta_{min} - \epsilon \to \theta_{min} + n\epsilon$$
$$\theta_{max} + \epsilon \to \theta_{max} - n\epsilon$$

(52)

where $n$ represents the learning rate and $n < N/2$. This Markov chain is aperiodic and irreducible.

At equilibrium, the Markov chain converges to a stationary or invariant probability distribution on the angles, $\pi(\theta_i)$, which satisfies the global balance equations. The invariant distribution can be calculated as the largest left eigenvalue/eigenvector of the probability transition matrix $q$, in which the $(i,j)$ entry is the probability of transitioning from state $i$ to state $j$.

For concreteness, we illustrate within an example. For the states

$$\theta_i = \{\theta_{min} - \epsilon, \theta_{min}, \theta_{min} + \epsilon, \theta_{min} + 2\epsilon = \theta_{max} - 2\epsilon, \theta_{max} - \epsilon, \theta_{max}, \theta_{max} + \epsilon\}$$

(53)

and transition matrix

$$q = \begin{bmatrix} 0 & 0 & 1 & 0 & 0 & 0 & 0 \\ \frac{1}{2} & 0 & \frac{1}{2} & 0 & 0 & 0 & 0 \\ 0 & \frac{1}{2} & 0 & \frac{1}{2} & 0 & 0 & 0 \\ 0 & 0 & \frac{1}{2} & 0 & \frac{1}{2} & 0 & 0 \\ 0 & 0 & 0 & \frac{1}{2} & 0 & \frac{1}{2} & 0 \\ 0 & 0 & 0 & 0 & \frac{1}{2} & 0 & \frac{1}{2} \\ 0 & 0 & 0 & 0 & 1 & 0 & 0 \end{bmatrix}$$

(54)

the invariant distribution is

$$\pi = \begin{bmatrix} \frac{1}{18} & \frac{1}{9} & \frac{2}{9} & \frac{2}{9} & \frac{2}{9} & \frac{1}{9} & \frac{1}{18} \end{bmatrix}$$

(55)

We see explicitly that values of $\theta$ outside the range $[\theta_{max}, \theta_{min}]$ are discouraged ($\pi_1 = \pi_7 = \frac{1}{18}$) and that $\theta_{min}$ and $\theta_{max}$ are less likely than other values of $\theta$ which lie strictly within the range ($\frac{1}{9}$ vs $\frac{2}{9}$). In practice, there may be other transitions from $\theta$ outside the range to $\theta$ within the range, which is controlled by the learning rate. For example, replacing $q_{12} = q_{13} = q_{76} = q_{75} = \frac{1}{2}$ and $q_{1i} = q_{7i} = 0$ for all other $i$ in Eq. (54) leads to

$$\pi \approx \begin{bmatrix} 0.053 & 0.105 & 0.210 & 0.263 & 0.210 & 0.105 & 0.053 \end{bmatrix}$$

(56)

in which there is even more asymmetry among the states within the range.

*Parallelization of differentiable rotamer sampling with improved memory usage*
To take advantage of the parallel computing potential of GPU and TPU backends, we followed an approach like that of AlQuraishi's parallelized natural extension reference frame (pNeRF) algorithm (*39, 40*); however, our method generalizes to all rotational degrees of freedom.

We split protein chain into $N_f$ fragments, splitting at the peptide bond between the C1 carbon of residue $i$ and the N2 nitrogen of residue $i + 1$. We recorded the positions of atoms in each segment, $x_k^f$, where $f$ labels the fragment and $k$ labels the number of the atom. For the N-terminal of fragment $f$, we appended the positions of four atoms near the peptide bond (C1=O from the C-terminal residue of fragment $f - 1$ and N2-C$\alpha$ from the N-terminal residue of fragment $f$) and appended those four positions to the $x_k^f$; we selected the same four atoms from the C-terminal of fragment $f$ (C1=O from the C-terminal residue of fragment $f$ and N2-C$\alpha$ from the N-terminal residue of fragment $f + 1$). Next, we applied our original rotamer sampling method on each fragment independently. We then aligned the C-terminal of $f$ to the N-terminal of $f + 1$, thus joining the fragments together. Finally, we applied dihedral rotations using the Rodrigues' formula as discussed previously to the peptide bonds at which we initially split the protein chain, thus recovering the $N_f - 1$ degrees of freedom lost upon splitting the chain into $N_f$ pieces.

In addition to allowing for parallelization, this approach reduced memory usage. In our original dihedral sampling method, instead of transforming each position separately for each dihedral angle, we transformed all positions in the protein simultaneously, and then masked those positions which were not part of the associated connected component $CC(e_m)$.

Assuming that PyTorch's JIT compiler was not able to optimize the resulting computational graph, this approach could be memory-intensive and computationally inefficient, with roughly $N_{dih} \times N_{atoms}$ 3D matrix-vector multiplications and $2 \times N_{dih} \times N_{atoms}$ 3D vector additions. In backpropagation, each of the intermediate matrices must be used, resulting in $N_{dih} \times N_{atoms} \times 3$ floating point numbers being stored (a copy of the position matrix for each dihedral angle). In the parallel approach, ignoring the extra positions appended to each fragment, the number of dihedrals per fragment is roughly $\frac{N_{dih}}{N_f}$, and the number of positions per fragment is similarly $\frac{N_{atoms}}{N_f}$. The total number of floating point operations is therefore $N_f \times \left(\frac{N_{dih}}{N_f} \times \frac{N_{atoms}}{N_f}\right) = \frac{N_{dih} N_{atoms}}{N_f}$, so we have reduced the raw number of operations by a factor of $N_f$. The analogous calculation for memory usage shows a reduction by a factor of $N_f$. Assuming PyTorch launches and runs all $N_f$ kernels simultaneously on a GPU, we can achieve a speed-up over the original dihedral sampler approach of $N_f^2$ for large proteins, while using $\frac{1}{N_f}$ times as much memory.

Our parallelized version of differentiable rotamer sampling is summarized in Algorithm 6: DiffRotParallel. For practical purposes, lines 1-21 in Algorithm 6 only need to be executed once, and the rotamer sampling from lines 22 to the end may be placed in a separate function.

*Alignment of point clouds*
We used two methods to align point clouds. The first was the well-known Kabsch algorithm (*41*), and the second was through an alternative approach due to a quaternionic derivation by Coutsias et al. (*42*). We chose to employ the second method, since it reduced to a largest magnitude eigenvalue/eigenvector problem.

Given two sets of positions $x$ and $y$, represented by $N \times 3$ matrices $x_{ij}$ and $y_{ij}$, with $i$ labeling the particle number and $j$ labeling the 3D components, the goal is to find the best fit rotation matrix and translation which transform the positions of the particles in $x$ to the corresponding positions in $y$.

We first center the two sets of points, by averaging the 3D components over all particles

$$\bar{x}_j = \frac{1}{N}\sum_i x_{ij}$$
$$\bar{y}_j = \frac{1}{N}\sum_i y_{ij}$$
$$X_{ij} = x_{ij} - \bar{x}_j$$
$$Y_{ij} = y_{ij} - \bar{y}_j$$

(57)

We then calculate the covariance matrix

$$H_{ij} = \sum_k X_{ki} Y_{kj}$$

(58)

In the Kabsch method, we calculate the singular value decomposition

$$H = U\Sigma V^T$$

(59)

Since the Kabsch method may result in improper rotations, we correct for changes in basis orientation by calculating

$$d = \text{sign}(\det(VU^T))$$

(60)

and the optimal rotation matrix is expressed as

$$R = V \begin{bmatrix} 1 & 0 & 0 \\ 0 & 1 & 0 \\ 0 & 0 & d \end{bmatrix} U^T$$

(61)

In the quaternionic approach, the quaternion associated with the optimal rotation in 3D is the eigenvector $[q_0, q_1, q_2, q_3]$ associated with the largest magnitude eigenvalue of the traceless, symmetric matrix

$$F = \begin{bmatrix} H_{11}+H_{22}+H_{33} & H_{23}-H_{32} & H_{31}-H_{13} & H_{12}-H_{21} \\ H_{23}-H_{32} & H_{11}-H_{22}-H_{33} & H_{12}+H_{21} & H_{13}+H_{31} \\ H_{31}-H_{13} & H_{12}+H_{21} & -H_{11}+H_{22}-H_{33} & H_{23}+H_{32} \\ H_{12}-H_{21} & H_{13}+H_{31} & H_{23}+H_{32} & -H_{11}-H_{22}-H_{33} \end{bmatrix}$$

(62)

To convert the quaternion $q = (q_0, q_1, q_2, q_3)$ to a rotation matrix, we compute

$$R = \begin{bmatrix} q_0^2 + q_1^2 - q_2^2 - q_3^2 & 2(q_1q_2 - q_0q_3) & 2(q_1q_2 + q_0q_2) \\ 2(q_1q_2 + q_0q_3) & q_0^2 - q_1^2 + q_2^2 - q_3^2 & 2(q_2q_3 - q_0q_1) \\ 2(q_1q_3 - q_0q_2) & 2(q_2q_3 + q_0q_1) & q_0^2 - q_1^2 - q_2^2 + q_3^2 \end{bmatrix}$$

(63)

In the quaternionic approach, the sign of the eigenvalue determines the orientation of the rotation. We assumed that the ideal proper rotation for our uses always has the largest magnitude eigenvalue.

The quaternion approach is summarized in Algorithm 4: QuaternionAlignment.

*Differentiable largest eigenvalue and associated eigenvector of a square matrix*
Since linear algebra operations like `torch.linalg.eigh` for Hermitian matrices solve for all eigenvalues of the matrix, we used a modified version of the power iteration algorithm to determine the largest eigenvalue and associated eigenvector. Additionally, this method has the advantage of being implemented in terms of pure PyTorch operations (matrix multiplication and floating point division), which is convenient for PyTorch backends such as Apple's Metal Performance Shaders which do not yet have complete coverage of linear algebra operations.

Symmetric matrices are diagonalizable, have real eigenvalues, and have an orthonormal eigenbasis. Given such a matrix, for example the $4 \times 4$ matrix $A$ we may write

$$A = S \Lambda S^T$$

(64)

where $\Lambda = \text{diag}(\lambda_1, \lambda_2, \cdots, \lambda_n)$ is a diagonal matrix of eigenvalues and $S$ is an orthonormal matrix such that $S^T S = I$ and the $i^{th}$ column of $S$ is a normalized eigenvector $v_i$ corresponding to $\lambda_i$. Then for any matrix power $k$, we may write

$$A^k = (S \Lambda S^T)(S \Lambda S^T) \cdots (S \Lambda S^T) = S \Lambda^k S^T$$

(65)

Since $\Lambda$ is diagonal, its powers are also diagonal. Carrying out the matrix multiplications,

$$A^k = \sum_i v_i \lambda_i^k v_i^T = \lambda_{max}^k \sum_i \left(\frac{\lambda_i}{\lambda_{max}}\right)^k v_i v_i^T$$

(66)

$v_i v_i^T$ is the outer product of $v_i$ with itself:

$$v_i v_i^T = \begin{bmatrix} v_{i1} \\ v_{i2} \\ \vdots \\ v_{in} \end{bmatrix} \begin{bmatrix} v_{i1} & v_{i2} & \cdots & v_{in} \end{bmatrix} = \begin{bmatrix} v_{i1}(v_{i1}) & v_{i1}(v_{i2}) & \cdots & v_{i1}(v_{in}) \\ v_{i2}(v_{i1}) & v_{i2}(v_{i2}) & \cdots & v_{i2}(v_{in}) \\ \vdots & \vdots & \ddots & \vdots \\ v_{in}(v_{i1}) & v_{in}(v_{i2}) & \cdots & v_{in}(v_{in}) \end{bmatrix}$$

(67)

When $k$ is large, the relative contribution of each eigenvalue and its eigenvectors to $A^k$ decreases in comparison to that of the maximum eigenvalue/eigenvector pair. To achieve large powers of $A$, we may repeatedly square $A$, so that the relative contributions of the smaller eigenvalues decreases exponentially. Thus with large $k$, (in practice we chose $k = 20$) we have

$$A^k \approx \lambda_{max}^k v_{max} v_{max}^T \qquad (68)$$

In practice, when we square $A$, we need to normalize by the largest magnitude element (or any other matrix norm) to prevent $A$ from drifting off to infinity or 0 due to the factor of $\lambda_{max}^k$.

We may directly read off the $v_i$ from the outer product because each row and each column of $v_{max} v_{max}^T$ is some multiple of $v_{max}$. To recover the eigenvalue $\lambda_{max}$, we simply apply the original matrix using the eigenvector equation $A v_{max} = \lambda_{max} v_{max}$ and compute

$$\lambda_{max} \approx \frac{1}{n} \sum_i \frac{(A v_{max})_i}{(v_{max})_i} \qquad (69)$$

The power iteration through repeated squaring method is summarized in Algorithm 5: PowerIterWithSquaring. Since we only required this method for a few $4 \times 4$ covariance matrices, it contributed a negligible amount of overhead to the overall algorithm. It is novel to this work, and can be thought of as a modified version of the power iteration algorithm (*43*, *44*).

*Estimation of entropy*
For estimation of distribution entropy of our Boltzmann generators, we used a method tailored for multivariate circular distributions (*45*), which attempts to mitigate correlations among the angles. The metric for two sets of angular samples $\phi, \psi$ was defined as the arclength on the unit circle for each angle

$$d^2(\phi, \psi) = \sum_i \left[\pi - |\pi - |\phi_i - \psi_i||\right]^2 \qquad (70)$$

Given a batch of samples $\Phi = [\phi, \psi, \cdots, \omega]$, we then computed the nearest neighbor for each sample in the batch $\Phi' = [\phi', \psi', \cdots, \omega']$ according to the metric $d$, and estimated the entropy of each sample as (Eq. 17 in the original manuscript (*45*), with first nearest neighbors corresponding to $m = 1$)

$$H_\phi = \log d(\phi, \phi') \qquad (71)$$

and averaged over the entire set of samples $\phi \in \Phi$. Since we were only using the entropy to provide approximate gradients to the Boltzmann generator, we ignored all constants which corrected for bias to the numerical value of the entropy in the original formula.

*Estimation of temperature of Boltzmann-generated samples*
We assumed that the energies of Boltzmann-generated samples follow an exponential distribution

$$E(x) \sim e^{-\beta x}$$
(72)

To reduce the influence of outliers on our estimate, we used the median energy of the sampled states as an estimator of the inverse temperature $\beta$. It is known analytically (*46*) that

$$E_{median} = \frac{\log 2}{\beta}$$
(73)

*Additional useful identities*
To aid in debugging of rotation matrices, we made use of the formula for the rotation angle and axis of an arbitrary $3 \times 3$ rotation matrix $R$.

$$\theta = \cos^{-1}\left(\frac{\operatorname{tr} R - 1}{2}\right)$$
(74)

$$\text{axis} = \frac{1}{2 \sin \theta} [R_{32} - R_{23} \quad R_{13} - R_{31} \quad R_{21} - R_{12}]^T$$
(75)

where $\operatorname{tr}$ is the trace. These identities are direct consequences of Rodrigues' rotation formula.

*Learning rate tuning*
To ensure convergent training, we found that a learning rate of $\gamma = 10^{-5}$ was necessary to prevent divergence during training due to large gradients. To arrive at this value of $\gamma$, we backpropagated gradients from the force field and entropic terms separately to the angles output by the neural network (i.e., only backpropagation only through the rotameric sampling and entropy estimation portions of the computational graph). We monitored these gradients over the course of a training session and found that angular gradients had a magnitude on the order of $10^3$. Therefore, parameters are tuned at a rate of approximately $10^3 \cdot \gamma$, per epoch, ignoring the momentum contributions in Adam (*47*). With the choice $\gamma = 10^{-5}$, we expect parameter tuning at a rate of roughly 0.01 per epoch; along with the choice of gradient clipping at 10.0, we enforced a large dynamic range of gradients with magnitudes under 0.01 to 10.0.

*Neural network architecture*
We used simple feedforward neural networks as our Boltzmann generators, with 32 latent variables, 10 layers of 128 hidden units, dropout with a rate of 0.3 (*25*), residual connections (*48*) between every other hidden layer, and LeakyReLU (*49, 50*) activations with a coefficient of 0.3. The final output layer dimension equal to the total number of dihedrals we wished to sample.

*Traditional molecular dynamics*
We performed traditional MD using the AMBER 14 force field (*2*) with generalized Born implicit solvent (*18–22*) in OpenMM (*32*). We used a Langevin thermostat at 310 K, friction coefficient of $91 \text{ ps}^{-1}$, and timestep of $1 \text{ fs}$. We simulated $10^6$ timesteps (1 ns trajectory), and recorded positions of all atoms every 1,000 timesteps.

*Order parameters*
We calculated the root mean square deviation (averaged over the entire structure) and root mean square fluctuation (averaged over a trajectory per alpha-carbon), with respect to reference structure positions $x_{i,0}$.

$$\text{RMSD}(\{x_i\}) = \sqrt{\frac{1}{N_{atoms}} \sum_i |x_i - x_{i,0}|^2}$$

$$\text{RMSF}(\{x_i\})_j = \sqrt{\frac{1}{N_{timesteps/samples}} \sum_t |x_j(t) - x_{j,0}|^2}$$

*Implementation details*
All algorithms were implemented in Python 3.10 with PyTorch 1.13 (*16*). We loaded molecular topology and geometry using OpenMM 7.7 (*32*). We used PDBFixer 1.8.2 (*32*) to fix errors in PDB files and model hydrogens. We used NetworkX 2.8.4 (*51*) for all graph algorithms. We used a batch size of 8, and the Adam optimizer (*47*), with learning rate $\gamma = 10^{-5}$, momentum coefficients $\beta_1 = 0.9$, $\beta_2 = 0.999$, and machine tolerance offset $\epsilon = 10^{-8}$. We also used a loss weight of $\gamma_{weight} = 10^{-2}$ for weight decay regularization.

For the molecular force field, we used AMBER 14 parameters (*2*) as provided by OpenMM (*32*), and the implicit generalized Born solvent (GBn2) (*18, 20, 22, 35*). Other packages used include NumPy 1.23.4 (*52, 53*) and Pandas 1.5.2 (*54*) for data organization, Matplotlib 3.6.2 (*55*) for plotting, PyMol 2.5.0 (*56*) and Mol* (*57*) for macromolecule visualization, and SciPy 1.9.3 (*58*). Training of models was performed in float32 accuracy on the CPU only of an M1 Max MacBook Pro with 64 GB RAM. Benchmarking was also performed on a single NVIDIA Tesla T4 GPU with 16 GB RAM on a Linux system. For reproducibility, we also set the random seed for PyTorch to 0.

To optimize the speed of dihedral angle application, we used static data structures and used PyTorch's just-in-time compilation on the dihedral sampling function, as well as the generative neural networks. This technique allows PyTorch to optimize numerical array operations through fusion of kernels and memory locality. As a result, we observed that the primary bottleneck for our pipeline was the force field computation and the transfer of positional data between PyTorch and OpenMM.

**Supplemental Tables**

Table S1: Units of measurement and specific numerical values used to represent molecules and their properties in OpenMM and PyTorch. This table may be used to convert the numerical outputs of the computations to predicted measurements.

| Physical quantity | Physical units | Numerical value during computation |
|---|---|---|
| distance $(d_{ij})$ | nanometer (nm) | 1 |
| energy $(E, \epsilon_{ij})$ | kilojoule / mole (kJ mol$^{-1}$) | 1 |
| electric charge $(q)$ | elementary charge $(e = 1.609 \times 10^{-19}$ C) | 1 |
| Avogadro's constant $(N_A)$ | unitless | $6.022 \times 10^{23}$ |
| Ideal gas constant $(R)$ | kilojoule / (mole Kelvin) (kJ mol$^{-1}$ K$^{-1}$) | $8.314 \times 10^{-3}$ |
| Coulomb constant $\left(\frac{1}{4\pi\epsilon_0}\right)$ | kJ nm mol$^{-1}$ $e^{-2}$ | 138.9354576 |
| Solute (protein) relative electric permittivity $(\epsilon_{solute})$ | unitless | 1 |
| Solvent (water) relative electric permittivity $(\epsilon_{solvent})$ | unitless | 78.5 |